# On Galois-Division Multiple Access Systems: Figures of Merit and Performance Evaluation

J. P. C. L. Miranda, H. M. de Oliveira

CODEC – Communications Research Group
Departamento de Eletrônica e Sistemas – CTG – UFPE
C.P. 7800, 50711-970, Recife – PE, Brazil
e-mail: jump@elogica.com.br , hmo@ufpe.br

*Abstract - A new approach to multiple access based on finite field transforms is investigated. These schemes, termed Galois-Division Multiple Access (GDMA), offer compact bandwidth requirements. A new digital transform, the Finite Field Hartley Transform (FFHT) requires to deal with fields of characteristic p, p ≠ 2. A binary-to-p-ary (p ≠ 2) mapping based on the opportunistic secondary channel is introduced. This allows the use of GDMA in conjunction with available digital systems. The performance of GDMA is also evaluated.*

## 1. INTRODUCTION

The interest in wireless communications has growth rapidly in the past two decades. The total number of wireless subscribers is expected to double over the next five years, while the total wireless minutes-of-use will increase even faster (Figure 1). Bandwidth is therefore becoming such an important good.

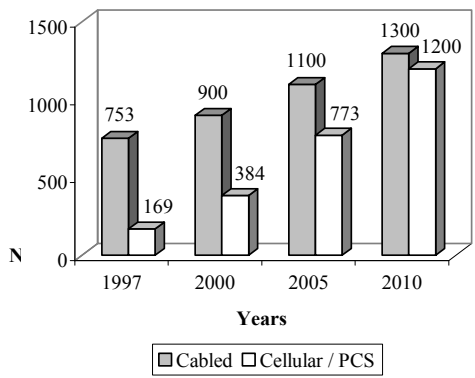

*Fig.1: Source: Northern Business Information, World Bank, ITU, LT.*

The Direct Sequence Code Division Multiple Access (DS-CDMA) is becoming the most popular multiple access scheme [17]. Recently, a new approach named Galois-Division Multiple Access (GDMA) was introduced [5] as an efficient-bandwidth multiple access scheme for band-limited channels. GDMA focussing on properties of the Finite Field Fourier Transform (FFFT) was already presented in [6].

## 2. AN OVERVIEW ON GDMA SYSTEMS

GDMA systems are digital multiple access schemes, which employ spread spectrum techniques based on finite field transforms, such as the FFFT introduced by Pollard [16]. Information strings coming from users are combined by means of alphabet expansion. The users' sequences $\underline{v}$ are defined over a finite field GF($p$). Thus, the multiplex (mux) involves an expanded signal set whose symbols are defined over the extension field, GF($p^m$). That is, given a signal $\underline{v}$ over the ground field GF($p$), we deal with the Galois domain by considering the spectrum $\underline{V}$ over an extension field GF($p^m$), which exactly corresponds to the FFFT of the signal $\underline{v}$ [1]. Once the mux can be carried out by a FFFT, the demux corresponds exactly to its inverse, i.e., the FFFT$^{-1}$.

Assume that each symbol over GF(p) has a duration $T$ seconds. An N-user GDMA (N-GDMA) can be designed over GF($p^m$) where $N \mid p^m - 1$. Clearly, a constraint on the number of users is made. This happens because the length of FFFTs is related to the order of the elements constituting its kernel. The spread signal is then generated by a componentwise multiplication between the input signal $\underline{v} = (v_0, v_1, ..., v_{N-1})$, $v_i \in$ GF($p$), and a Galois-Carrier $\underline{c} = (c_0, c_1, ..., c_{N-1})$, $c_i \in$ GF($p^m$). These carriers should be interpreted as multilevel pseudo-noise spreading sequences [13].

A key feature of this multilevel coding-division multiplex (multilevel DS-CDMA) is its ability to improve bandwidth requirements. Besides the FFFT, a new finite field transform [3] named Finite Field Hartley Transform (FFHT) can be used. It is a new finite field version of the integral transform introduced by R. V. L. Hartley [9], [2]. The FFHT components belong to the group GI($q$), defined in the sequel.

**Definition 1**: G($q$) = { $\xi = \alpha + j\beta;\ \alpha, \beta \in$ GF($q$)}, $q = p^r$, $r$ being a positive integer, $p$ being an odd prime for which $j^2 = -1$ is a quadratic non-residue in GF($q$), is the set of Gaussian integers over GF($q$).   ∎

Let $\otimes$ denote the Cartesian product. The set G($q$) together with the operations $\oplus$ and $*$ defined bellow is a field.

$\oplus$ : G($q$) $\otimes$ G($q$) $\to$ G($q$)
$(\alpha_1 + j\beta_1, \alpha_2 + j\beta_2) \to (\alpha_1 + j\beta_1) \oplus (\alpha_2 + j\beta_2) =$
$= (\alpha_1 + \alpha_2) + j(\beta_1 + \beta_2).$

$* : G(q) \otimes G(q) \to G(q)$
$(\alpha_1 + j\beta_1, \alpha_2 + j\beta_2) \to (\alpha_1 + j\beta_1) * (\alpha_2 + j\beta_2)$
$= (\alpha_1\alpha_2 - \beta_1\beta_2) + j(\alpha_1\beta_2 + \alpha_2\beta_1)$.

It can be shown [11] that the algebraic structure $GI(q) = <G(q), \oplus, *>$ is a field. In fact, GI(q) is isomorphic to $GF(q^2)$.

It is well known that finite field transforms contains some redundancy [16], [1]. The GDMA bandwidth saving lays on its ability to purge this redundancy. The procedure of transmitting only leaders of cyclotomic cosets, from now on denoted by cyclotomic compression, leads to more compact bandwidth requirements. Two other points should be mentioned. GDMA systems are based on transforms for which there exist fast algorithms [18]. They are also convenient from the hardware point of view since implementation can carried out by a digital signal processor (DSP).

**Definition 2**: The bandwidth compactness factor is defined as $\gamma_{cc} = N / \nu$, where $N$ is the number of users and $\nu$ is the number of cyclotomic cosets associated with an FFFT / FFHT spectrum. ∎

It should be remarked that, when cyclotomic compression is performed, the clock driving FFFT symbols is $N / \nu$ times faster than the input baud rate.

The next step is to answer questions like: Can the bandwidth compactness factor be increased without limit? Which are the trade-offs involved? Does cyclotomic compression give rise to any performance loss? An attempt to provide satisfactory answers to these questions can be found in the remainder of this paper.

### 3. AN UPPER BOUND ON THE COMPACTNESS FACTOR $\gamma_{cc}$

System trade-offs are fundamental to all digital communications designs. Some goals to be achieved by the designer are clearly in conflict with the others. There are several constraints and theoretical limitations that necessitate the trading-off of any one requirement with each of the others. In this section, one of the most famous theoretical constraints will be applied: The Shannon capacity theorem.

**Proposition 1**: Let SNR be the average signal-to-noise ratio. The bandwidth compactness factor is upper bounded by the following expression:

$$\gamma_{cc} \leq \log_p(1 + SNR). \quad (1)$$

**Proof**: The Shannon-Hartley capacity of a Gaussian channel is given by:
$$C = W_{GDMA} \log_2(1 + SNR) \text{ bits / s}.$$

Reliable information transmissions over additive white Gaussian channels (AWGN) at a rate $R$ are possible, provided that $R \leq C$. Arbitrarily small error rates can be achieved by using a sufficiently complicated coding scheme. Conversely, it is not possible to find out a code which achieve an arbitrarily small error rate for $R > C$. Therefore:

$$R_{GDMA} \leq W_{GDMA} \log_2(1 + SNR) \text{ bits / s}. \quad (2)$$

Supposing an N-GDMA, it is straightforward to verify that the global data rate is given by

$$R_{GDMA} = \frac{N \log_2 p}{2T} \text{ bits / s}, \quad (3)$$

and the required bandwidth according to [5] is:

$$W_{GDMA} = \frac{N}{2T} \frac{1}{\gamma_{cc}} \text{ Hz}. \quad (4)$$

The proof follows by substituting (3) and (4) into (2). ∎

Expression (1) shows that more compact bandwidth can be achieved by means of a larger alphabet extension, i.e. larger $\gamma_{cc}$, which depends uniquely on the signal-to-noise ratio.

### 4. BINARY TO $P$-ARY MAPPING: THE TRANSCODIFICATION

Digital communication systems usually deal with binary data strings. Thus, users are represented by binary sources. The GDMA concept, however, demands $p$-ary sequences. In fact, GDMA user's signal set is defined over GF($p$). Note that if $p = 2$ no compatibility problem will be experienced. On the other hand, the FFHT is constrained to fields of characteristic $p \neq 2$ [3]. Then, some binary-to-$p$-ary mapping is required when practical implementations are considered.

In earlier applications the opportunistic secondary channel was conceived as a mean to create an opportunity to transmit additional side information [8]. More generally, it can be used for increasing the information bit rate by a small fractional number, i.e., as a way to introduce an additional rate in generalised constellations [7], [4]. A binary-to-$p$-ary mapping can be carried out by applying the opportunistic secondary channel reasoning, as showed in Examples 1 and 2. This approach is referred herein to as an alphabet transcodification. The supplied mapping is one-to-one and there exist thus no forbidden sequences.

Example 1: GF(2)-to-GF(7) mapping (see Table I).

Table I: Binary-to-7-ary Mapping.

| GF(7) elements | Code A | Code A' |
|---|---|---|
| 0 | 000 | 000 |
| $\alpha^0 = 1$ | 001 | 010 |
| $\alpha^1 = 3$ | 11 | 11 |
| $\alpha^2 = 2$ | 010 | 100 |
| $\alpha^3 = 6$ | 110 | 101 |
| $\alpha^4 = 4$ | 100 | 001 |
| $\alpha^5 = 5$ | 101 | 011 |

Galois field points corresponding to Code A are sketched in Figure 2.

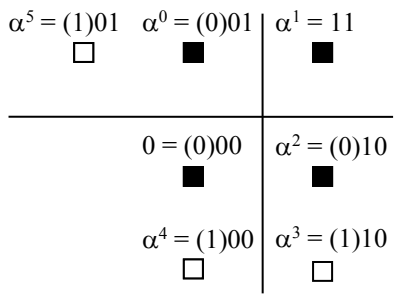

*Fig.2: 7-point generalised assignment using the opportunistic secondary channel. The basic constellation is the square 4-point one with shaded points. 3 more points are appended to 3 of its points and bear the same 2-bit label. A third opportunistic bit (in brackets) is appended to tell if the corresponding point belongs or not to the basic constellation.*

If instantaneous codes are preferred, the opportunistic bit can be appended at the end of each code word. The new code (code A´) is also shown in Table I. The following mapping illustrates this approach:

1 0 **1** 1 0 **0** 1 1 0 **1** 1 1 **1** 0 ... $\leftrightarrow \alpha^3 \ \alpha^2 \ \alpha \ \alpha^5 \ \alpha$ ...

Opportunistic bits are boldface typed.

An intermediate step is required to map extension fields or isomorphisms such as GI($p$). Elements of GF($p^m$) are viewed as m-dimensional vectors, while elements of GI($p$) can be viewed as two-dimensional vectors. In both cases coefficients lie in the ground field GF($p$). The decimal value of $p$-ary codewords when translated into binary representation yields the map of interest. Extra bits demanded by alphabet transcodification correspond exactly to opportunistic bits. Typically, basic signal processing functions take place over GF(2). Thus, the alphabet transcodification presented here plays a very important role at GDMA design, making such schemes compatible with available digital systems. Furthermore, the binary-to-$p$-ary mapping also become possible the implementation of GDMA systems based on the FFHT. However, it is important to mention that practical implementation is rather difficult due to the asynchronous nature of signalling.

Example 2: GF(2)-to-GI(3) mapping (see Table II).

Table II: GI(3) Complex Field Elements Map.

| GI(3) elements | 3-ary Code | Code B |
|---|---|---|
| 0 | 00 | 0000 |
| $\xi^0 = 1$ | 10 | 011 |
| $\xi^1 = 1 + j$ | 11 | 100 |
| $\xi^2 = 2j$ | 02 | 010 |
| $\xi^3 = 1 + 2j$ | 12 | 101 |
| $\xi^4 = 2$ | 20 | 110 |
| $\xi^5 = 2 + 2j$ | 22 | 1000 |
| $\xi^6 = j$ | 01 | 001 |
| $\xi^7 = 2 + j$ | 21 | 111 |

Galois symbol assignment is sketched in Figure 3.

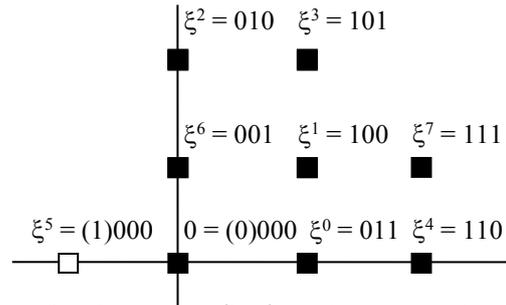

*Fig.3: 9-point generalised assignment using the opportunistic secondary channel.*

## 5. ERROR-CORRECTION CAPABILITY

GDMA error-correction ability was left to be investigated in [5]. Instead of eliminating redundancy (by means of cyclotomic compression), one could transmit all of coefficients in order to introduce some error-correction capability. In fact, this statement was based on observing that FFFT valid spectrum sequences generates a multilevel block code.

**Proposition 2**: The FFFT valid spectra form a multilevel block code ($N$, $2^N$, 1) over GF($2^m$) with minimum distance equals to unit.

**Proof**: Let $\{V_K^r\}$ and $\{V_K^s\}$ be FFFT valid spectra. Then, $\forall \ \beta, \chi \in$ GF(2),

$$\beta\{V_K^r\} + \chi\{V_K^s\} = \beta\sum_{i=0}^{N-1} v_i^r \alpha^{ik} + \chi\sum_{i=0}^{N-1} v_i^s \alpha^{ik}$$

$$= \sum_{i=0}^{N-1} (\beta v_i^r + \chi v_i^s)\alpha^{ik}$$

$$= \sum_{i=0}^{N-1} v_i^t \alpha^{ik} .$$

Therefore $\beta\{V_K^r\} + \chi\{V_K^s\}$ is also an FFFT valid spectrum. Minimum distance computation considers the minimisation of Hamming distance between the spectra (codewords) $\{V_K^r\}$ and $\{V_K^s\}$. That is:

$$d = \underset{r \neq s}{\text{Min}} \ d_H(\{V_K^r\},\{V_K^s\})$$

$$= \underset{r \neq s}{\text{Min}} \ W_H\{V_K^r - V_K^s\},$$

where $W_H$ denotes the Hamming weight of a sequence. This can be computed in term of the Hamming weight of components, $w_H$, according to

$$= \underset{r \neq s}{\text{Min}} \sum_{k=0}^{N-1} w_H(V_K^r - V_K^s)$$

$$= \underset{r \neq s}{\text{Min}} \sum_{k=0}^{N-1} w_H\left[\sum_{i=0}^{N-1}(v_i^r - v_i^s)\alpha^{ik}\right].$$

Defining a new binary sequence $\{v_i^t\}=\{v_i^r\}-\{v_i^s\}$, it follows that

$$d = \underset{t}{\text{Min}} \sum_{k=0}^{N-1} w_H\left(\sum_{i=0}^{N-1} v_i^t \alpha^{ik}\right).$$

Now, let $\{v_i^t\}$ = (1, 1, 1, ..., 1). Using correlation properties of Galois-Fourier carriers [6], it is easy to find that the corresponding Galois spectrum $\{V_K^t\}$ is (1, 0, 0, ..., 0), which is always a valid spectrum. Then, d = 1 = $W_H$ (1, 0, 0, ..., 0). ∎

Although containing some redundancy, full spectrum GDMA signals have null error-correction capability, as Proposition 2 shows. If error-correction capability is required, system's designer should perform cyclotomic compression and then apply a Reed-Solomon code [12].

In the next section the performance of GDMA is analysed in terms of the symbol error rate (SER).

## 6. ASYMPTOTIC BEHAVIOUR OF SER

This section' goal is to establish what is lost (or won) as the number $N$ of GDMA users increases. We will also be able to derive a reasonable approximation for the asymptotic behaviour of SER, under cyclotomic compression effects.

Previous investigations on N-GDMA systems, $N = p^m - 1$, suggested that the number of constellation points required for modulation should be $M = N + 1 = p^m$. From the practical viewpoint, this is a severe constraint. That is, given a channel (and thus a SNR), a high number of users will require modulations with a great number of states, which are difficult to implement and often have unsatisfactory performance. Fortunately, using the opportunistic secondary channel again, N-GDMA systems can afford supporting arbitrary digital modulation techniques, no matter the number of users.

**Definition 3**: For an N-GDMA system the $h$ parameter is defined as the (average) number of modulation symbols required to convey a single Galois symbol. ∎

**Proposition 3**: Given an $M$-point constellation, $M$ a power of two, and $p^m = 2^s + W$, with $W$ being the number of points appended to the basic assignment, the average number of transmitted modulation symbols per GDMA frame is given by

$$hN = \frac{RN}{\log_2 M} \text{ modulation symbols/frame,}$$

where $R$ is the average number of information bits per transmitted Galois symbol.

**Proof**: The concept of an opportunistic secondary channel results in transmitting an amount of s bits per Galois symbol when the direct channel is used. Conversely, $s + 1$ bits (per Galois symbol) are transmitted when the opportunistic secondary channel is selected. Let $P_{dir}$ and $P_{opp}$ be the probability of selecting the direct channel and the opportunistic channel, respectively. The average transmission rate (in information bits per Galois symbol) is given by:

$$R = P_{dir}m + P_{opp}(m+1) \text{ bits/Galois symbol.} \quad (5)$$

Supposing equally likely symbols, $P_{dir}=2^s/p^m$ and $P_{opp}=W/p^m$. The digital modulation conveys ($\log_2 M$) bits/modulation symbol, yielding an average rate of $h=R/\log_2 M$ modulation symbol per Galois symbol. A transmitted frame at each GDMA input symbol interval corresponds to the finite field transform vector (FFFT or FFHT) of input data. Since the transform blocklength is $N$ (equal to the number of users) the result follows. ∎

Example 3: Suppose an 8-GDMA system based on the FFHT. Transmitted Galois symbols are defined over GI(3), as showed in Table II. Assuming the source bits to be independent and equally likely, the next three bits to be transmitted are examined in order to determine whether the opportunistic channel is used:

$$P_{dir} = P(\{\xi^0, \xi^1, \xi^2, \xi^3, \xi^4, \xi^5, \xi^6, \xi^7\}) = 7/8$$
$$P_{opp} = P(0) = 1/8.$$

From (5) follows that:

$$R = 7/8.(3) + 1/8.(4) = 3.125 \text{ bits per GI(3) symbol}.$$

The average number of transmitted modulation symbols per 8-GDMA frame can be found in Table III, for different digital modulation techniques.

Table III: $h$-Parameter for Different Modulations.

| h values (modulation symbols per Galois symbols) | | | | | |
|---|---|---|---|---|---|
| BPSK | QPSK | 8-PSK | 16-QAM | 32-QAM | 64-QAM |
| 3.125 | 1.562 | 1.042 | 0.781 | 0.625 | 0.521 |

**Proposition 4**: Let an N-GDMA system employing arbitrary digital modulations over an additive white noise channel. An upper bound on the frame error rate is [15]:

$$P_{E,N} \leq hN.P_{E,1},$$

where $P_{E,1}$ is the symbol error-rate for one user.

Proposition 4 states that cyclotomic compression does not give rise to any performance loss. In the worst case, system performance will be left unchanged; e.g., the case of 8-GDMA systems based on FFHT [15].

It is often convenient to specify the system performance by the probability of a bit error or bit error rate (BER), even when decisions are made on the basis of symbols ($\log_2 M > 1$). Then, the definition of performance almost universally accepted is the BER.

## 7. BER ESTIMATION

Simulating a communication system often involves analysing its response to the noise inherent in real-world components. Such analysis allows an investigation of the system's response and helps designing the system. There are a number of ways to arrive at an estimate of the BER [10], each with its own advantages and disadvantages.

Example 4: Consider a primitive polynomial over GF(2), say $p(x) = x^4 + x + 1$. Let $\alpha$ be a primitive element of GF(16) and, therefore $N = \text{ord}(\alpha) = 15$. The extension field GF(16) can be generated as usual [14]. If the designer considers the FFFT, the Galois-Fourier carriers over GF(16) will be $\{\alpha^{ik}\}$, $k = 0, 1,..., N – 1$.

In this example the number of FFFT cyclotomic cosets is $\nu_F = 5$. They are: C0 = (0), C1 = (1, 2, 4, 8), C3 = (3, 6, 12, 9), C5 = (5, 10) and C7 = (7, 14, 13, 11). The cyclotomic compression is obtained by transmitting only the leader of each cyclotomic coset. Therefore, selecting cyclotomic leaders, the transmitted vector will be $\underline{V}_{comp} = (V_0, V_1, V_3, V_5, V_7)$. Indeed, $\gamma_{cc}=15/5=3$.

The retrieval of missing spectral components is carried out using valid spectra properties. In this case: $V_k^2 = V_{2k \,(\text{mod}\,15)}$. Despreading precisely corresponds to perform an inverse FFFT of length 15. The recovered vector is then multiplied by a sequence with all components identical to $1 / N \,(\text{mod}\, p)$. Carriers $\{\alpha^{-ik}\}$, $k = 0, 1, ..., N–1$, have components equals to the multiplicative inverse of the Galois-Fourier carrier components over GF(16).

The results presented in this section were obtained using the most general of BER estimation techniques: the Monte Carlo method [10]. For each BER point estimated were observed 50 million bits. Figure 4 plots show the BER estimation for 15-GDMA systems based on the FFFT over AWGN channels for various modulation schemes. Both full-spectrum (FS) transmissions and compressed (CC) transmissions were considered.

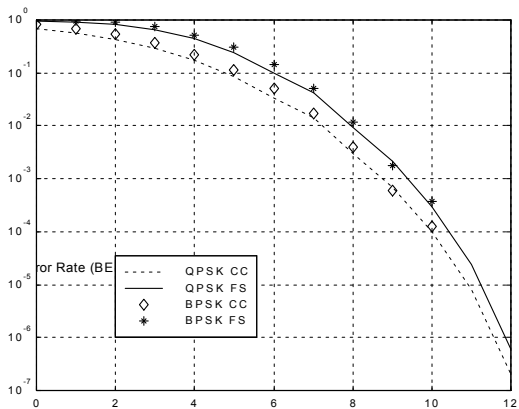

(a)

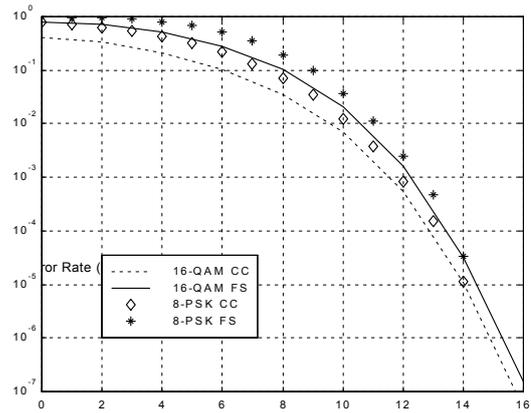

(b)

*Fig.4: BER estimation for 15-GDMA systems based on the FFFT over AWGN channels (Monte Carlo method). (a) BPSK and QPSK. (b) 8-PSK and 16-QAM.*

## 8. CONCLUDING REMARKS

For the case of GDMA systems, the following trade-offs might be made:

- Bandwidth gain can be achieved by taking higher alphabet extensions (i.e., using best $\gamma_{cc}$ results). Limitations on bandwidth savings were examined.

- The SNR per bit required to achieve a given BER can be reduced by decreasing N. This can be done by carrying out a cyclotomic compression or dealing with shortened-GDMA systems [6].

- Actually, cyclotomic compression gives rise to some performance gain. For evaluated systems a maximum gain of one decibel was obtained regarding full spectrum transmissions.

An approach to uncouple GDMA technique from the digital modulation was supplied. Its foundation consists in conveying Galois symbols (from the finite field) through unmatched constellation symbols of an arbitrary digital modulation. The performance of GDMA was evaluated by simulation for different modulation schemes. Regarding practical implementations, two new valuable contributions were given. The opportunistic secondary channel was offered as a suitable tool for a binary-to-$p$-ary mapping and for modulating GDMA signals using arbitrary digital modulation techniques. It should be noticed that both applications are quite different from the original approach.

It was also shown in this paper that, despite their intrinsic redundancy, GDMA systems based on the FFFT have null error-correction capacity. Combined multiplex and Reed-Solomon coding is left to be investigated.

ACKNOWLEDGEMENTS

The first author would like to thank The Brazilian National Council for Research and Development (CNPq) for its partial financial support and also Dr. E. Fontana for his interest in this research. He also thanks L. G. Caldeira for constructive discussions about BER estimation techniques.